# Design of the Reverse Logistics System for Medical Waste Recycling Part I: System Architecture and Disposal Site Selection Algorithm

Chaozhong Xue[#], Yongqi Dong[#], *Student Member, IEEE*, Jiaqi Liu, Yijun Liao, and Lingbo Li

*Abstract*— With social progress and the development of modern medical technology, the amount of medical waste generated is increasing dramatically. The problem of medical waste recycling and treatment has gradually drawn concerns from the whole society. The sudden outbreak of the COVID-19 epidemic further brought new challenges. To tackle the challenges, this study proposes a reverse logistics system architecture with three modules, i.e., medical waste classification & monitoring module, temporary storage & disposal site (disposal site for short) selection module, as well as route optimization module. This overall solution design won the Grand Prize of the "YUNFENG CUP" China National Contest on Green Supply and Reverse Logistics Design ranking 1[st]. This paper focuses on the description of architectural design and the module on site selection. Specifically, regarding system architecture, a framework diagram is provided, together with brief descriptions of the three proposed modules and a case study under the COVID-19 epidemic with the customized model. Regarding the disposal site selection module, a multi-objective optimization model is developed, and considering different types of waste collection sites (i.e., prioritized large collection sites and common collection sites), a hierarchical solution method is developed employing linear programming and K-means clustering algorithms sequentially. The proposed site selection method is verified with a case study using real-world data, and compared with the baseline, it can immensely reduce the daily operational costs and working time. Limited by length, detailed descriptions of the whole system as well as the remaining medical waste classification & monitoring module and route optimization module can be found at https://shorturl.at/cdY59.

## I. Introduction

In recent decades, medical waste generation is rising dramatically. If the large amount of medical waste is not properly managed and recycled, it might cause serious risks for both human beings and the environment [1]. Thus, the public is gradually increasing their concerns about the management of healthcare/medical waste on a global basis [2], especially after the outbreak of the Coronavirus Disease 2019 (COVID-19) epidemic [3]–[7]. To cope with the arising challenges, there is a must for designing a reliable and effective medical waste reverse logistics system that is greatly beneficial to the whole society.

Regarding waste reverse logistics system development, most studies focus on network design [4], [5], [7]–[12]. For example, in [4], a novel multi-objective & multi-period mixed integer program was developed for the reverse logistics network design, especially in epidemic outbreaks. It mainly aims to tackle the exponentially increased medical waste within a very short period by optimizing temporary facilities localization as well as transportation strategies. The designed model was applied to a case study concerning the outbreak of COVID-19 in Wuhan, China. In [9], the medical waste reverse logistics network design problem was tackled from the perspective of medical materials producers, with a mixed integer linear programming (LP) model developed for handling the returned medical waste. In [10], to minimize the overall costs and the risks associated with the operations of infectious medical waste, an LP model with three objective functions is developed, plus multiple functions are designed to calculate the amount of generated waste according to the parameters tuned from the epidemic outbreak. There are seldom studies that try to design an integrated reverse logistics system integrating site selection and route optimization for the recycling medical waste transportation while considering the classification & monitoring of the waste. This study aims to fill such research gaps with an advanced architectural design. Specifically, this study tackles the optimization of reverse logistics transportation in two phases, i.e., the temporary storage & disposal site (disposal site for short) selection phase and the route optimization phase.

Regarding the disposal site selection, there are seldom studies considering the second-level temporary storage & disposal centers. Most of the available studies are about the general facility location selection [13], [14]. This study attempts to delve into the direction of setting up second-level temporary storage & disposal centers and the corresponding optimization for site selection for such centers.

## II. Reverse Logistics System Architecture Design

To tackle the aforementioned challenges and fill the research gaps in designing the reverse logistics system, while considering the current situation and problems regarding recycling medical wastes in China, this study proposes and designs a system architecture with three modules, i.e., waste classification & monitoring module, temporary storage & disposal site selection module, as well as route optimization module. The system architecture and the relationships between each module are illustrated in Figure 1, while the

[#] These authors contributed equally to this work and should be considered as co-first authors.

Chaozhong Xue is with Department of Traffic Engineering and Key Laboratory of Road and Traffic Engineering, Tongji University, Shanghai, 200092, China. (e-mail: 1750700@tongji.edu.cn).

Yongqi Dong is with Delft University of Technology, Delft, 2628 CN, the Netherlands, he is also with University of California, Berkeley, Berkeley, CA 94720, USA. (e-mail: yongqidong@berkeley.edu).

Jiaqi Liu is with Department of Traffic Engineering and Key Laboratory of Road and Traffic Engineering, Tongji University, Shanghai, 200092, China. (corresponding author, phone: +86 18019113619; e-mail: liujiaqi13@tongji.edu.cn).

Yijun Liao is with Department of Traffic Engineering and Key Laboratory of Road and Traffic Engineering, Tongji University, Shanghai, 200092, China. (e-mail: 1851021@tongji.edu.cn).

Lingbo Li is with Department of Traffic Engineering and Key Laboratory of Road and Traffic Engineering, Tongji University, Shanghai, 200092, China. (e-mail: 2233430@tongji.edu.cn).

remaining parts of this section will briefly introduce the motivation to set up these modules as well as their main functions. Detailed descriptions for each module will be provided in the following sections.

*A. Medical Waste Classification and Monitoring Module*

There are many defects in the current medical waste disposal management system, e.g., lacking effective traceability management tools, and blind spots in the disposal process, resulting in unscrupulous black industrial chains driven by the interest of profit. This study proposes an "Internet+" based medical waste traceability management platform using image recognition, RFID, and other technical assistance, to integrate medical waste management into the information management channel, implementing effective, real-time, visual monitoring of the entire process of medical waste disposal. This will provide logistics companies, medical institutions, and relevant government departments with all-weather 24/7 real-time monitoring without dead ends which technically cuts off the black industrial chains, thus eliminating the hidden dangers in the reuse process. To be specific, this module involves three entities (i.e., relevant government departments, hospitals, and logistics companies.) and provides five functions, i.e., intelligent weighing management, tracking management using BeiDou Navigation, tracking management using RFID, medical waste collection & inlet/outlet management, and classification evaluation management. The corresponding data flow among the three main entities together with their corresponding roles are illustrated in the top-level data flow diagram in Figure 2. More detailed descriptions are provided in the supplementary materials at https://shorturl.at/cdY59.

*B. Temporary Storage & Disposal Site Selection Module*

Due to the wide and scattered sources of medical waste in the recycling process, the daily waste generation of small and medium-sized private medical institutions is small. While, to date, the layout of their temporary storage and disposal centers is usually unreasonable and cannot reach the economic scale of setting up such disposal operation facilities. So it is necessary to set up temporary storage and disposal sites, and the selection and layout of such sites need to be upgraded and optimized. This is identified by this study as one of the key parts of constructing the medical waste reverse logistics system. To tackle the problems of the low daily production of waste from low-level medical units (e.g., community hospitals, outpatient clinics, and other small clinics), and to upgrade the unreasonable layout of temporary storage and centralized disposal sites, in this module, this study first divides medical waste generation sites into prioritized large collection sites (e.g., Primary hospitals and above), and common collection sites (e.g., outpatient clinics). Then, a multi-objective optimization model is developed considering the number of the selected temporary storage & disposal sites, and their corresponding recycling amounts, together with the costs of transfer and recycling. Finally, a hierarchical solution method is developed employing linear programming and K-means clustering algorithms sequentially to solve the developed model.

*C. Medical Waste Recycling Route Optimization Module*

With the settled temporary storage & disposal sites, the optimization of medical waste recycling transportation also needs to consider route planning. It is necessary to take into account both the cost of logistics recycling companies and the low level of risk required by the government. In order to find the equilibrium point between the two objectives of minimizing cost and maximizing safety, this study constructs a mathematical model for recycling route planning. Firstly, this study divides the cost of the medical waste recycling reverse logistics into two parts, i.e., vehicle transportation cost and risk cost. Then, the model is constructed with minimizing the weighted joint costs as the objective. The Analytic Hierarchy Process (AHP) method is adopted to determine the model parameter weights. Finally, this study develops a customized Genetic Algorithm (GA) and employs it to solve the model. As verified by a case study, this module design can help to minimize transportation risks and reduce the total cost of collection and transportation, while maximizing the efficiency of collection and transportation, and meeting the satisfaction of medical institutions.

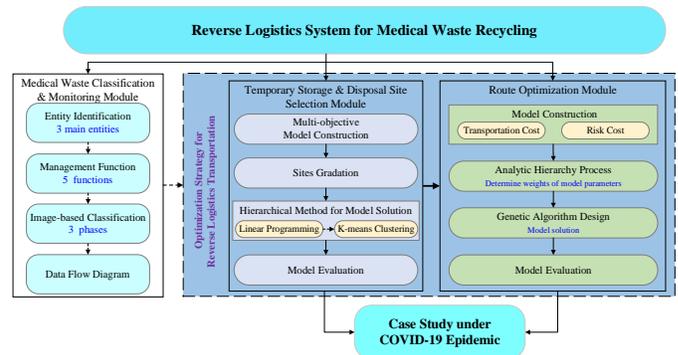

Figure 1. Illustration of the proposed reverse logistics system architecture

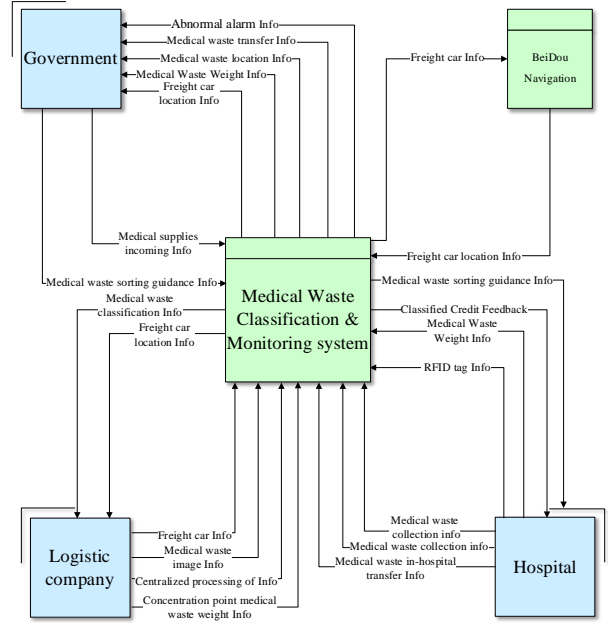

Figure 2. The top-level data flow diagram among the entities

*D. Customized Model and Case Study under COVID-19 Epidemic*

Since the outbreak of the COVID-19 epidemic, the volume of medical waste has increased exponentially across countries, plus the waste during the epidemic is highly infectious, which has put forward higher requirements for medical waste recycling, transportation, and treatment. In the epidemic, medical waste recycling companies are faced with a series of challenges such as huge task loads, tight schedules, and high standards, and are in need to come up with efficient and feasible solutions for medical waste recycling. Based on the developed modules of site selection and route optimization, together with relevant developed models, this study further customizes an urban medical waste recycling method by setting the time window and penalty cost considering the "day and night" emergency transportation mode (which is introduced in detail in the supplementary material) and the cold-chain transportation requirements. A customized GA algorithm is adopted to solve the upgraded model. As verified by the case study using empirical data from Dalian city, it is observed the upgraded model can obviously improve the performance under the situation of setting up special emergency lines for responding to the COVID-19 epidemic.

Limited by length, detailed descriptions of the route optimization module and the customized model for the case study under the COVID-19 epidemic are distributed in another paper which serves as supplementary material and can be found at https://shorturl.at/cdY59.

## III. TEMPORARY STORAGE & DISPOSAL SITE SELECTION MODULE DESIGN

### A. Problems Identification and Module Framework Design

In the process of recycling medical waste, there are many source sites of medical waste with relatively scattered distributions, among which the daily output of medical waste in low-level medical units (e.g., community hospitals, and outpatient clinics) is small. So regardless of collecting wastes from all the medical units, it is necessary to set up temporary storage and disposal sites. The current unreasonable layout of temporary storage and disposal sites cannot reach the expected economic scale. So it is necessary to optimize the temporary storage and disposal site selection which will serve as one of the major parts of reverse logistics transportation. However, this is usually neglected or downplayed in previous studies (e.g., [4], [5], [12]). To tackle the aforementioned problems and fill the research gaps, this study tackles the optimization of reverse logistics transportation in two phases, i.e., temporary storage & disposal site selection phase and route optimization phase. In the developed specific site selection module, the study first divides medical waste generation sites into prioritized large collection sites and common collection sites, then develops a multi-objective optimization model, and finally establishes a hierarchical solution method employing linear programming and K-means clustering algorithms sequentially to solve the developed model. Figure 3 illustrates the framework design of this temporary storage & disposal site selection module.

### B. Optimizaiton Model Construction

#### a) Basic assumptions

1) It is assumed that there is no limit to the capacity of medical waste at all levels of collection points.
2) The recycling process will involve the establishment of temporary disposal centers at large collection points, and

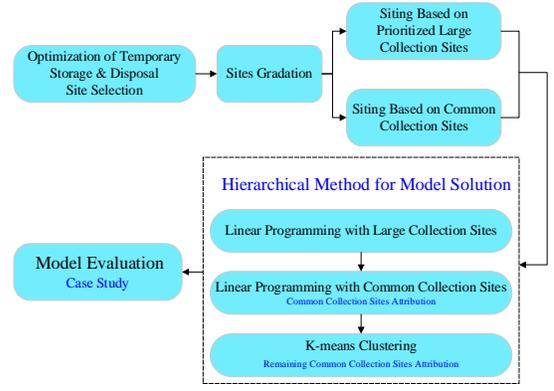

Figure 3. Illustration of disposal site selection module.

TABLE I. VARIABLE DEFINITION

| Variables | Definition |
|---|---|
| $M_1$ | A set of common waste collection sites that $|N_1| = m$ |
| $M_2$ | The set of first-level large collection sites and temporary storage and disposal centers, $|M_2| = c$ |
| $q_i$ | Amount of medical waste generated at common collection sites |
| $q_j$ | The amount of medical waste generated by the first-level waste collection site $j$ ( i.e., temporary storage and disposal center) itself |
| $Q_i$ | Maximum capacity limit of waste collection site $i$ |
| $f_j$ | The fixed construction cost of setting up temporary storage and disposal center |
| $b$ | Costs of medical waste disposal in temporary disposal centers |
| $t$ | Medical waste unit transfer cost |
| $a_1$ | The unit subsidy of medical waste treatment in the temporary disposal center, which is linearly related to the volume of medical waste transferred to the temporary disposal center |
| $a_2$ | Unit subsidy for common recycling sites transfers, with subsidy cost linearly related to distance |
| $N$ | A set of common collection sites that are not covered by large collection sites |
| $S$ | The set of common collection sites that are covered by large collection sites |
| $L$ | Distance threshold from the common collection site $i$ to large collection site $j$ |

the temporary disposal centers will require certain additional costs, including fixed construction costs for building new centers or upgrading the small common sites, and unit disposal costs, while the government will give a subsidy to set up temporary disposal centers.

3) The transfer of medical waste from the common waste collection sites to the temporary disposal center will also incur a transfer cost, and again the government will provide a transfer subsidy to encourage this initiative.

#### b) Variable declarations

The variable declarations are illustrated in TABLE I.

#### c) General objective function

If the distance from common collection site $i$ to large collection site $j$ is exceeding L, this specific common collection site will not be covered, denoted as $N = \{i | \forall d_{ij} > L, i \in M_1, j \in M_2\}$. $d_{ij}$ is the distance from common collection site $i$ to large collection site $j$. If the number of elements in $N$ is $n$, then the set of common collection sites that are covered by large collection sites

$S = M_l\text{-}N$, with $|S| = m\text{-}n$. And then the objective function can be written as:

$$\min\ Z_1 = \sum_{j=1}^{c} w_j(f_j + b\sum_{i=1}^{m-n} H_{ij}q_i + bq_j - a_1\sum_{i=1}^{m-n} H_{ij}q_i) + \sum_{i=1}^{m-n}\sum_{j=1}^{c}(t - a_2)q_i d_{ij}H_{ij} \quad (1a)$$

s.t.

$$\sum_{j=1}^{c} H_{ij} = 1 \quad (1b)$$
$$d_{ij}H_{ij} \leq L \quad (1c)$$
$$\sum_{i=1}^{m-n} H_{ij} - Nw_j \leq 0 \quad (1d)$$
$$b > a_1 \quad (1e)$$
$$t > a_2 \quad (1f)$$
$$H_{ij} \in (0,1) \quad (1g)$$
$$w_j \in (0,1) \quad (1h)$$

where regarding the decision variables $w_j$ and $H_{ij}$,

$$w_j = \begin{cases} 1, & \text{if the large collection site } j \text{ is selected} \\ 0, & \text{other} \end{cases}$$

$$H_{ij} = \begin{cases} 1, & \text{if the wastes at the common site } i \text{ is} \\ & \text{transferred to the large site } j \\ 0, & \text{other} \end{cases}$$

The objective function (1a) indicates that the target cost consists of the fixed cost of the temporary disposal center and the treatment cost of medical waste. And the government subsidy will reduce part of the cost accordingly. To goal is to find the decision variables so that the total cost is minimized. Constraint (1b) indicates that the common collection site is transferred to only one temporary storage & disposal center. Constraint (1c) indicates the coverage distance constraint, once this range is exceeded, the cost becomes infinite. Constraint (1d) indicates the relationship between $H_{ij}$ and $w_j$. Constraint (1e) indicates that the treatment cost of the temporary storage & disposal center is greater than the subsidy. Constraint (1f) indicates that the transfer cost is greater than the unit subsidy. Constraints (1g) and (1h) are 0-1 variable constraints.

*d) Objective function design when selecting sites from the common collection sites that are not covered by the large collection sites as the temporary storage & disposal sites*

To select sites from the common collection sites that are not covered by the large collection sites (i.e., sites in $N$) as the temporary storage & disposal sites, this study introduces the decision variables $F_i$ and $U_{ij}$, where

$$F_i = \begin{cases} 1, & \text{if the common site is selected} \\ 0, & \text{other} \end{cases}$$

$$U_{ij} = \begin{cases} 1, & \text{if the wastes at the common site } i \text{ is} \\ & \text{transferred to the large site } j \\ 0, & \text{other} \end{cases}$$

Then the modified objective function can be written as:

$$\min\ Z_2 = \sum_{i=1}^{n} F_i(f_i + b\sum_{i=1}^{n} U_{ij}q_i + bq_i - a_1\sum_{i=1}^{m-n} U_{ij}q_i) + \sum_{i=1}^{n}\sum_{j=1}^{n}(t - a_2)q_i d_{ij}U_{ij} \quad (2a)$$

s.t.

$$d_{ij}U_{ij} \leq L \quad (2b)$$
$$q_j + \sum_{i=1}^{n} U_{ij}q_i \leq Q_i \quad (2c)$$
$$\sum_{j=1}^{n} U_{ij} = 1 \quad (2d)$$
$$b > a_1 \quad (2e)$$
$$t > a_2 \quad (2f)$$
$$U_{ij} \in (0,1) \quad (2g)$$
$$F_i \in (0,1) \quad (2h)$$

The objective function (2a) indicates that the target total cost is composed of the fixed cost and the transfer cost from the other common collection sites to the ones selected as the temporary storage & disposal centers. The objective is to find decision variables $F_i$ and $U_{ij}$ to minimize the total cost. As for the constraints, constraint (2b) indicates that the transfer distance cannot be greater than the maximum transfer radius, i.e., the distance threshold L. Constraint (2c) indicates that the capacity of the common collection site selected as the temporary disposal center is constrained to be the sum of the medical waste generated by itself and the transferred volume from other sites. Constraint (2d) means that wastes at any collection point can only be transferred to at most one temporary disposal center. Constraint (2e) means that the treatment cost of the temporary disposal center is greater than the subsidy. Constraint (2f) means that the transfer cost is greater than the unit subsidy. Constraints (2g) and (2h) are 0-1 variable constraints.

*C. Hierarchical Method for Model Solution*

To solve the aforementioned model, this study develops a hierarchical method with three layers employing LP and K-means clustering algorithms as illustrated in Figure 3.

Firstly, all **Primary** hospitals or above are identified as large collection sites for medical waste recycling. One can get the decision variables $w_j$ and $H_{ij}$ by solving the minimization problem defined by (1a) using the LP algorithm [15]. However, considering the large amount of medical waste produced by large collection sites every day, plus the fact that there are not too many large collection sites, so, in the case study (shown later), the results suggested that all large collection sites can be set up as temporary storage & disposal centers. This serves as the first layer of the model solution.

Then, regarding the objective function (2a) with its linear corresponding constraints, this study employs the LP algorithm to solve it. In practice, both the two LP problems, (1a) and (2a), are solved by utilizing LINGO [16]. After solving (1a) and (2a), with the determined decision variables $w_j$, $H_{ij}$, $F_i$ and $U_{ij}$, all common sites within distance L of all the selected temporary storage & disposal center sites are assigned a center site. This serves as the second layer of the model solution method.

Finally, for those remaining common sites that are still not assigned to any centers, this study adopts K-means clustering [17] to find a suitable temporary storage and disposal center for them. The number of additional center sites is determined by tuning the hyperparameter $K$ referring to the clustering results and elbow method. Once $K$ is settled, all the remaining common sites are supposed to be assigned to their corresponding cluster centers. However, considering the costs of setting up new sites, instead of using the cluster centers, this study suggests using the nearest common collection site as the center selection for each cluster. The detailed process of the K-means clustering algorithm is as follows:

**Step 1)** Randomly select $K$ sites as the clustering centers;

TABLE II.    ESTIMATION OF DAILY MEDICAL WASTE GENERATION

| Organization Type | Primary Hosptial and above | Community Hospital | Outpatient | Clinic |
|---|---|---|---|---|
| Amount of medical waste generation | 0.4 kg/day/bed | 15-20 kg/day | 15-20 kg/day | 1.5 kg/day |

TABLE III.    ESTIMATION OF MODEL PARAMETERS

| Variables | Definition | Estimated Value |
|---|---|---|
| $f_j$ | The fixed construction cost of the storage and disposal center | 3000 CNY |
| $b$ | Costs of medical waste disposal in temporary disposal centers | 3 CNY/kg |
| $t$ | Medical waste unit transfer cost | 2 CNY/kg |
| $a_1$ | The unit subsidy of medical waste treatment in the temporary disposal center, the subsidy is linearly related to the volume of medical waste transferred to the temporary disposal center | 1 CNY/kg |
| $a_2$ | Unit subsidy for common recycling sites transfers, with subsidy linearly related to distance | 0.5 CNY/kg |
| $L$ | Distance threshold from the common collection site $i$ to large collection site $j$ | 500 m |
| $Q_i$ | Maximum capacity limit of the collection sites | 1500 kg |

**Step 2)** Calculate the distance of each remaining site to each of the $K$ clustering centers, and then assign the site to its nearest clustering center, so that $K$ new clusters will be obtained.

**Step 3)** Recalculate and update the center of each cluster.

**Step 4)** Repeat the above steps 2~3 until the position of the cluster center no longer changes (within an acceptable threshold) or the iteration limit is reached.

The K-means clustering algorithm serves as the third layer of the proposed model solution method.

## IV. CASE STUDY

To verify the effectiveness of the proposed temporary storage & disposal site selection model, this paper carries out an instance analysis using real-world empirical data.

### A. Data Collection

In this study, the empirical data from Dalian, a city in northern China were selected to carry out the case study. An application was created connecting with the Baidu map open platform (API) to crawl the specific information of the target Dalian city area. Considering the complexity of data processing and solution scale, all medical units in two districts, Xigang District and Zhongshan District of Dalian City, were finally selected as the research target areas. There are a total of 112 medical units in the selected area, including 21 Primary (and above) hospitals and 91 small private hospitals and outpatient clinics. All the detailed locations of these 112 medical units were obtained with latitude and longitude thus it is possible to calculate the distance (in this study Manhattan distance is adopted) between them.

### B. Relevant Parameter and Parameter Assignments

Due to the varying levels of hospitals, the amount of medical waste generated varies. According to investigations and experience, the estimations of the amount of medical waste generated by different levels of hospitals are shown in TABLE II. Other parameters were determined by reviewing relevant literature and using surveys, and reasonable assessments for the relevant parameters are listed in TABLE III.

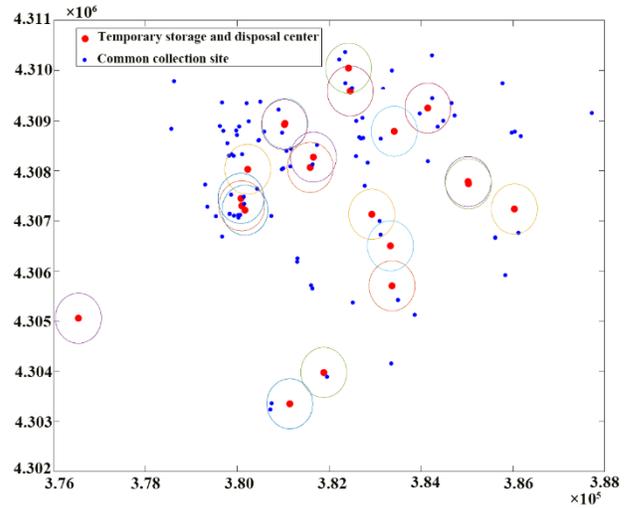

Figure 4.  Results demonstration after center siting based upon large sites and linear programming.

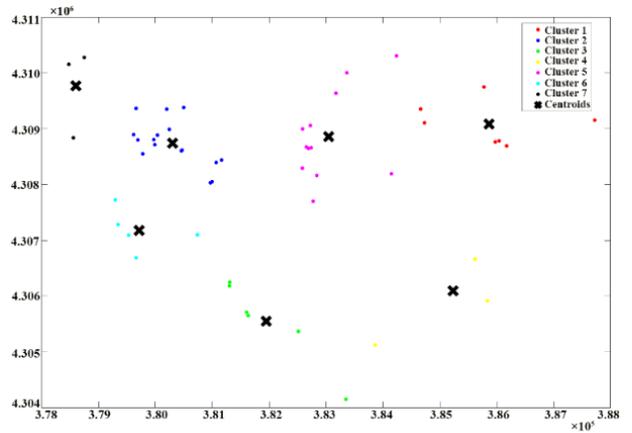

Figure 5.  Results of the K-means clustering.

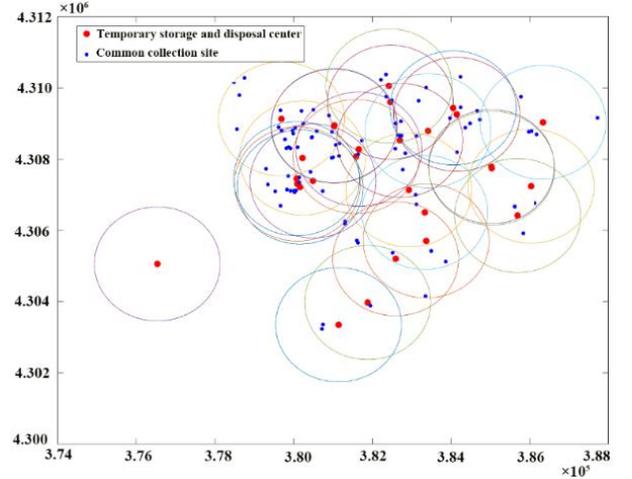

Figure 6. Final results of the temporary storage and disposal site selection.

### C. Model Solution

Firstly, after solving the corresponding minimization problem defined by (1a), in this case, all large collection sites are set up as temporary storage and disposal centers. While the common collection sites within L=500m of these selected storage and disposal centers (i.e., large collection sites in this case) are attributed to their corresponding centers using an LP algorithm. The result is demonstrated in Figure 4.

The next step is to select temporary storage and disposal centers and assign the attribution for all remaining unassigned sites by solving the corresponding minimization problem (2a), i.e., the 2nd layer of the model solution. Then, the final remaining sites' attribution together with new centers will be determined based on the clustering algorithm (the 3rd layer of the model solution). The common collection sites that are nearest to their cluster centers are the additional selected temporary disposal centers. And through tuning the parameters of $K$, it was finally determined that adding 7 more temporary centers would have the best improvement. Corresponding clustering results are shown in Figure 5. While the final results of the temporary storage & disposal sites selection module are illustrated in Figure 6. It is observed that all the sites are assigned with their corresponding centers with 28 centers in total. Please note, due to different settings of the coordinate limit and scale, the site distribution in Figure 4 ~ Figure 6 may look different, however, they are the same. One needs to check them carefully.

*D. Results Analysis and Evaluation*

Firstly, it is illustrated that the obtained 28 centers can cover all the sites within the selected study area. Since there is no information about the current temporary storage and disposal center sites, this study then seeks to compare the results with those without setting up the proposed second-level temporary storage and disposal center sites. From the estimated calculation, based on the objective function together with TABLE II, TABLE III, and relevant tables in the supplementary material at https://shorturl.at/cdY59, without the setting up of temporary storage and disposal center sites, the total daily working time for disposing the recycled medical waste (note this time does not include the transportation time between the sites) will be **1194** min, and the total daily maintenance cost for all sites will be **34212.9** CNY.

In comparison, after implementing the proposed temporary storage & disposal site selection method, the total daily working time for disposing the recycled medical waste will be **497** min, and the total daily maintenance cost for all sites will be **26928.9** CNY. Both of them are reduced dramatically, i.e., reduced at the level of **58.4%** and **20.8%** respectively.

The detailed process of the calculation for the total daily working time and maintenance costs is illustrated in the supplementary material at https://shorturl.at/cdY59.

## V. CONCLUSION

To handle the challenges regarding reverse logistics systems for medical waste recycling, especially those that arise during the COVID-19 epidemic, this study develops an integrated system architecture with three modules, i.e., medical waste classification & monitoring module, temporary storage & disposal site selection module, plus route optimization module. Limited by length, this paper is specifically focusing on the description of the proposed architectural design regarding the motivation, function, and benefits of each module. Also, it introduces the temporary storage & disposal site selection module in detail. Considering different types of waste collection sites (e.g., prioritized large collection sites and common collection sites), a multi-objective linear programming optimization model is proposed with corresponding hierarchical solution methods incorporating linear programming and K-means clustering algorithms sequentially to solve it. The proposed site selection method is verified with a case study using real-world data and compared with the baseline, the proposed method can immensely improve the performance reducing the daily operational costs and working time at a large margin. This site selection module is connected to the route optimization module which is described in detail in another paper available in the supplementary material at https://shorturl.at/cdY59.

REFERENCES

[1] S. Kargar, M. M. Paydar, and A. S. Safaei, "A reverse supply chain for medical waste: A case study in Babol healthcare sector," Waste Manag., vol. 113, pp. 197–209, 2020, doi: 10.1016/j.wasman.2020.05.052.

[2] E. Shinee, E. Gombojav, A. Nishimura, N. Hamajima, and K. Ito, "Healthcare waste management in the capital city of Mongolia," Waste Manag., 2008, doi: 10.1016/j.wasman.2006.12.022.

[3] E. Balci, S. Balci, and A. Sofuoglu, "Multi-purpose reverse logistics network design for medical waste management in a megacity: Istanbul, Turkey," Environ. Syst. Decis., vol. 42, no. 3, pp. 372–387, 2022, doi: 10.1007/s10669-022-09873-z.

[4] H. Yu, X. Sun, W. D. Solvang, and X. Zhao, "Reverse logistics network design for effective management of medical waste in epidemic outbreaks: Insights from the coronavirus disease 2019 (COVID-19) outbreak in Wuhan (China)," Int. J. Environ. Res. Public Health, 2020.

[5] K. Govindan, S. Nosrati-Abarghooee, M. M. Nasiri, and F. Jolai, "Green reverse logistics network design for medical waste management: A circular economy transition through case approach," J. Environ. Manage., vol. 322, no. July, p. 115888, 2022.

[6] M. Liu and T. Dai, "Optimization Strategy for Reverse Logistics Network of Medical Waste under COVID-19," Chinese Control Conf. CCC, vol. 2022-July, no. 71771120, pp. 1934–1939, 2022.

[7] A. R. P. Santos et al., "A Mixed Integer Linear Programming for COVID-19 Related Medical Waste Reverse Logistics Network Design," ACM Int. Conf. Proceeding Ser., pp. 473–477, 2022.

[8] X. Luo and W. Liao, "Collaborative Reverse Logistics Network for Infectious Medical Waste Management during the COVID-19 Outbreak," Int. J. Environ. Res. Public Health, vol. 19, no. 15, 2022.

[9] L. H. Shi, "A mixed integer linear programming for medical waste reverse logistics network design," 2009, doi: 10.1109/ICMSE.2009.5317680.

[10] S. Kargar, M. Pourmehdi, and M. M. Paydar, "Reverse logistics network design for medical waste management in the epidemic outbreak of the novel coronavirus (COVID-19)," Sci. Total Environ., vol. 746, p. 141183, 2020, doi: 10.1016/j.scitotenv.2020.141183.

[11] Z. Wang, L. Huang, and C. X. He, "A multi-objective and multi-period optimization model for urban healthcare waste's reverse logistics network design," J. Comb. Optim., 2021.

[12] X. Mei, H. Hao, Y. Sun, X. Wang, and Y. Zhou, "Optimization of medical waste recycling network considering disposal capacity bottlenecks under a novel coronavirus pneumonia outbreak," Environ. Sci. Pollut. Res., vol. 29, no. 53, pp. 79669–79687, 2022.

[13] C. Kahraman and B. Oztaysi, Supply Chain Management Under Fuzziness: Recent Developments and Techniques. 2014.

[14] Q. Shi, H. Ren, X. Ma, and Y. Xiao, "Site selection of construction waste recycling plant," J. Clean. Prod., 2019.

[15] G. B. Dantzig, Linear Programming 2: Theory and Extensions. 2002.

[16] L. Schrage, Optimization Modeling with LINGO. Lindo System, 2006.

[17] A. Hartigan and M. A. Wong, "A K-Means Clustering Algorithm," J. R. Stat. Soc., 1979.